# Influence of GaN substrate miscut on the XRD quantification of plastic relaxation in InGaN

J. Moneta[1], M. Kryśko[1], J. Z. Domagala[2], E. Grzanka[1], G. Muziol[1], M. Siekacz[1], M. Leszczyński[1] and J. Smalc-Koziorowska[1]

[1] Institute of High Pressure Physics, Polish Academy of Sciences, Sokolowska 29/37, 01-142 Warsaw, Poland

[2] Institute of Physics, Polish Academy of Sciences, Aleja Lotników 32/46, 02-668 Warsaw, Poland

Abstract

In the epitaxy of semiconducting materials, substrate miscut is introduced to improve the morphology of deposited layers. However, in the mismatched epitaxial system, the substrate miscut also changes the stress geometry in the deposited layer, thus influencing the relaxation processes. In this paper, we show that InGaN layers grown on misoriented (0001)-GaN substrates relax by preferential activation of certain glide planes for misfit dislocation formation. Substrate misorientation changes resolved shear stresses, affecting the distribution of misfit dislocations within each dislocation set. We demonstrate that this mechanism leads to an anisotropic strain as well as a tilt of the InGaN layer with respect to the GaN substrate. It appears that these phenomena are more pronounced in structures grown on substrates misoriented toward $\langle 11\bar{2}0 \rangle$ direction than corresponding structures with $\langle 1\bar{1}00 \rangle$ misorientation. We reveal that the lattice of partially relaxed InGaN has a triclinic deformation, thus requiring advanced XRD analysis. The presentation of just a single asymmetric reciprocal space map commonly practiced in the literature can lead to misleading information regarding the relaxation state of partially relaxed wurtzite structures.

## 1. Introduction

Misoriented substrates (also called miscut substrates) are typically used to promote the step-flow growth mode desired for the epitaxy of the highest quality layers. Misorientation is a configuration in which the crystallographic planes of the chosen substrate orientation are not exactly parallel to the surface, but are slightly tilted to expose surface steps (Fig. 1(a)). In the case of cubic and hexagonal heterostructures, it is a common practice to assume tetragonal lattice distortion (Fig. 1(b)) in strain analysis by X-ray Diffractometry (XRD) of epitaxial layers which are mismatched to the substrate [*1*]. However, substrate misorientation induces stress asymmetry in the heterostructure [*2*]. Then, the assumption of the tetragonal distortion is not accurate because triclinic lattice deformation appears. Crystallographic layer-to-substrate tilt is a common feature of many types of mismatched epitaxial layers, e.g., InGaAs/GaAs[*3,4*], InGaP/GaP[*5*], ZnSe/GaAs[*6*], CdTe/GaAs[*7*], GaAs/Si[*8*], SiGe/Si[*9-12*], BiFeO$_3$/SrTiO$_3$[*13*]. It is also known for III-nitrides heterostructures (AlN, GaN and InN and their alloys) [*14-17*]. The results of strain quantification are then affected by existing crystal deformations. Two effects can be distinguished: (i) a rigid layer-to-substrate tilt $\kappa$ and (ii) a deformation of the layer lattice, i.e., the inclination (tilt $\tau$) between the **c**-axis and the normal to the **c**-planes of the layer lattice. Both tilts may have a common origin and usually coexist. There are two main sources of the epilayer deformation: (i) coherent matching of the layer to the substrate at the substrate atomic steps and (ii) preferential misfit dislocation formation [*18*].

The corrections for the tilt of the layer relative to the substrate are usually included in the standard XRD strain analysis. However, the triclinic deformation is usually neglected for cubic and hexagonal materials. The influence of triclinic lattice deformation on XRD measurements has been reported for cubic heterostructures like AlGaAs/GaAs[*19*] and InGaAs/GaAs[*20*]. The triclinic lattice deformation has also been reported for coherent InGaN [*21*], but it is widely believed that hexagonal approximation with layer tilt correction is sufficient for XRD strain quantification of nitride layers.

The aim of this work is to investigate how the (0001) GaN substrate misorientation (azimuth and miscut angle) influences the plastic relaxation and structural properties of InGaN layers. Relaxed InGaN layers can be used as pseudo-substrates for the growth of In-rich InGaN-based devices.[*22-29*] For such application it becomes crucial to properly determine the strain state of relaxed templates. We show that the relaxation of InGaN layers deposited on misoriented substrates leads to anisotropic strain and that these InGaN layers exhibit a layer-to-substrate tilt that affects the final pseudo-substrate misorientation. The hexagonal symmetry approximation applied to partially relaxed wurtzite layers grown on the miscut substrates leads to large errors in the estimated degree of relaxation. It is then important to take into account the triclinic deformation of the InGaN unit cell when studying relaxed InGaN layers using X-ray diffraction for the proper determination of the state of relaxation. Our investigation focuses on the InGaN/GaN system, but the results can be directly applied to other compressively strained wurtzite layers like GaN or AlGaN layers grown on AlN substrates. Although, the strain relaxation mechanism by the introduction of misfit dislocations is not dominant in the case of tensile strained layers, it has also been reported for tensile strained AlGaN layers grown on GaN [*30*] and some conclusions drawn here can also be transferred to this material system.

## 2. Background
### 2.1. Crystal deformations of the coherently strained InGaN layers

Coherent matching of the layer to the misoriented substrate induces stress asymmetry and results in non-zero shear strain components acting in the coherent layer [*2*]. This leads to the deformation of the layer lattice (Fig. 1(a)): a deformation tilt $\tau$ and a layer-to-substrate tilt $\kappa_n$ called Nagai's tilt [*3*]. The tilt $\kappa_n$ describes the relationship between the layer lattice and the substrate lattice, while the tilt $\tau$ refers to the deformation of the layer lattice itself. Both effects have been reported for coherent InGaN layers with respect to the underlying GaN [*17,21*]. According to the elasticity theory, the deformation tilt $\tau$ can be estimated after Ref. [*21*] as: $\tau = |atan[\varepsilon_{yz}/(1 + \varepsilon_{yy})] + atan[\varepsilon_{yz}/(1 + \varepsilon_{zz})]|$. In the case of InGaN layer deposited on GaN substrates misoriented from [0001] direction, tilt $\tau$ is the angle between the [0001] axis of the layer and the normal to the (0001) plane of the layer (Fig. 1(a)). For example, the tilt $\tau$ of fully coherent $In_{0.2}Ga_{0.8}N$ with respect to 0.8° **m**-misoriented GaN is expected to be about 0.07°. Its direction is defined by the acting strain. For compressively strained InGaN on GaN the direction of the tilt $\tau$ is opposite to the miscut direction. While for tensile strained AlGaN on GaN the direction of the tilt $\tau$ is towards the miscut direction.[*16*]

Following the Nagai's model [*3*], the tilt $\kappa_n$ of the heteroepitaxial coherent layer with respect to the substrate can be estimated as: $\tan\kappa_n = -(h_{layer}-h_{substrate})/h_{substrate})\tan\xi$, where $h_{layer}$, $h_{substrate}$ denote the corresponding step heights and $\xi$ is a substrate misorientation angle. It does not depend on the layer thickness. In the case of InGaN layer deposited on GaN substrates misoriented from [0001] direction, tilt $\kappa_n$ is an inclination of the (0001) plane of the layer to the (0001) plane of the substrate (Fig. 1(a)). The steps of c/2 height are assumed in the Nagai's model, where $c_{layer}$, $c_{substrate}$ are corresponding out-of-plane lattice parameters. For example, the tilt $\kappa_n$ of fully coherent $In_{0.2}Ga_{0.8}N$ with respect to 0.8° **m**-misoriented GaN is expected to be about 0.016°. For the systems where $c_{layer}>c_{substrate}$, like in the case of InGaN/GaN heterostructures, the tilt $\kappa_n$ is opposite to the substrate miscut direction, i.e., it reduces the miscut angle of the overall structure. For tensile strained AlGaN/GaN the direction of the tilt $\kappa_n$ is towards the miscut direction adding to the miscut angle of the overall structure. The deviations from the Nagai's model can occur with step bunching. Various steps configurations influencing layer tilting have been reported for GaN on sapphire epitaxy [15]. Kryśko et al.[17] measured $\kappa_n$ tilt for coherent InGaN layers with different In content and deposited on GaN substrates with different miscut and reported values smaller by about 10% than predicted by the Nagai's model related to rough morphology of epitaxial InGaN.

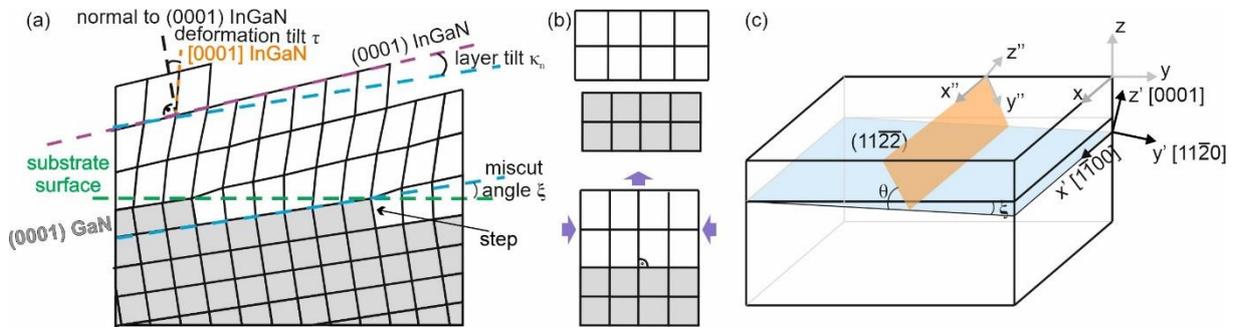

Fig. 1. (a) Schematic illustration of heteroepitaxial coherent layer grown on misoriented substrate with lattice parameters $c_{layer} > c_{substrate}$. Deformation of the layer lattice occurs due to coherent matching: the layer is tilted with respect to the substrate (tilt $\kappa_n$ between the **c**-planes of the layer and the substrate) and there is monoclinic distortion of the layer (tilt $\tau$ between the **c**-axis and the normal to the **c**-planes of the layer). (b) Schematic illustration of the tetragonal deformation of the layer with lattice parameters $a_{layer} > a_{substrate}$. (c) Scheme of the position of the $(112\bar{2}\bar{})$ slip plane in the wurtzite layer grown on the misoriented substrate. The coordinate systems used in the work are indicated: (x, y, z) related to the interface (surface), (x', y', z') related to the (0001) plane, (x", y", z") related to the slip plane. The (0001) basal plane and the $(112\bar{2}\bar{})$ slip plane are marked in blue and orange, respectively.

### 2.2. Introduction of misfit dislocations into a layer grown on misoriented substrate

In the epitaxial growth of materials with the same crystallographic structure, the lattice mismatch between the layer and the substrate leads to two phenomena: elastic deformation of the layer lattice and, under favourable conditions, additional formation of linear defects (misfit dislocations). The dislocations store energy and generate a strain field. The part of the mismatch accommodated by misfit dislocations can be expressed using elasticity theory. Coherent epitaxial layers are under biaxial stress:

$$[\sigma] = \begin{bmatrix} \sigma_{xx} & 0 & 0 \\ 0 & \sigma_{yy} & 0 \\ 0 & 0 & 0 \end{bmatrix}$$

(in the coordinate system related to the interface ((x, y, z), Fig. 1(c)). The wurtzite layer coherently grown on the (0001)-oriented substrate experiences an in-plane strain equal to the lattice mismatch:

$$\varepsilon_{xx} = \varepsilon_{yy} = f, \text{ where } f = \frac{a_{substrate} - a_{layer}}{a_{layer}}$$

and then an equibiaxial stress state:

$$\sigma_{xx} = \sigma_{yy} = \sigma = \frac{2(1-\nu)}{2-\nu} \varepsilon_{xx}. \text{ [31]}$$

The general case of strain and stress acting in wurtzite layers grown on substrates with surface normal tilted away from the [0001] direction was studied by Romanov et al.[2,32]. The authors have studied structures in non-polar and semi-polar orientations, i.e., they have analyzed the full range of inclination angle ξ from 0° to 90°. In our work, we analyze layers grown on substrates misoriented only up to ξ = 1.42°, so the stress is still close to the equibiaxial state.

The driving force behind the formation of misfit dislocations is the shear stress acting on the glide plane where movement occurs. Slip systems with higher resolved shear stresses experience a lower activation energy for dislocation nucleation and a higher glide velocity thus are therefore expected to dominate in the misfit dislocation formation process [4]. The glide geometry of the hexagonal crystal lattice is complex: three types of Burgers vectors ($\frac{1}{3}\langle 11\bar{2}0\rangle$, $\frac{1}{3}\langle 11\bar{2}3\rangle$, $\langle 0001\rangle$) and at least five different types of glide planes ((0001), {1$\bar{1}$00}, {11$\bar{2}$0}, {11$\bar{2}$2}, {1$\bar{1}$01}). For (0001)-oriented In$_x$Ga$_{1-x}$N films, the easy slip systems of the hexagonal lattice ($\frac{1}{3}\langle 11\bar{2}0\rangle$(0001), $\frac{1}{3}\langle 11\bar{2}0\rangle${1$\bar{1}$00}) are usually inactive due to the lack of resolved shear stresses.[33] Activation of the slip systems on pyramidal planes in such oriented layers ($\frac{1}{3}\langle 11\bar{2}3\rangle${11$\bar{2}$2}, $\frac{1}{3}\langle 11\bar{2}3\rangle${1$\bar{1}$01}) requires high energies and has been observed only in high quality layers, without other structural defects that can hinder the formation and propagation of 1/3<112,¯3> defects. [*33-35*] The observed misfit dislocations with $\frac{1}{3}\langle 112,\bar{3}\rangle$

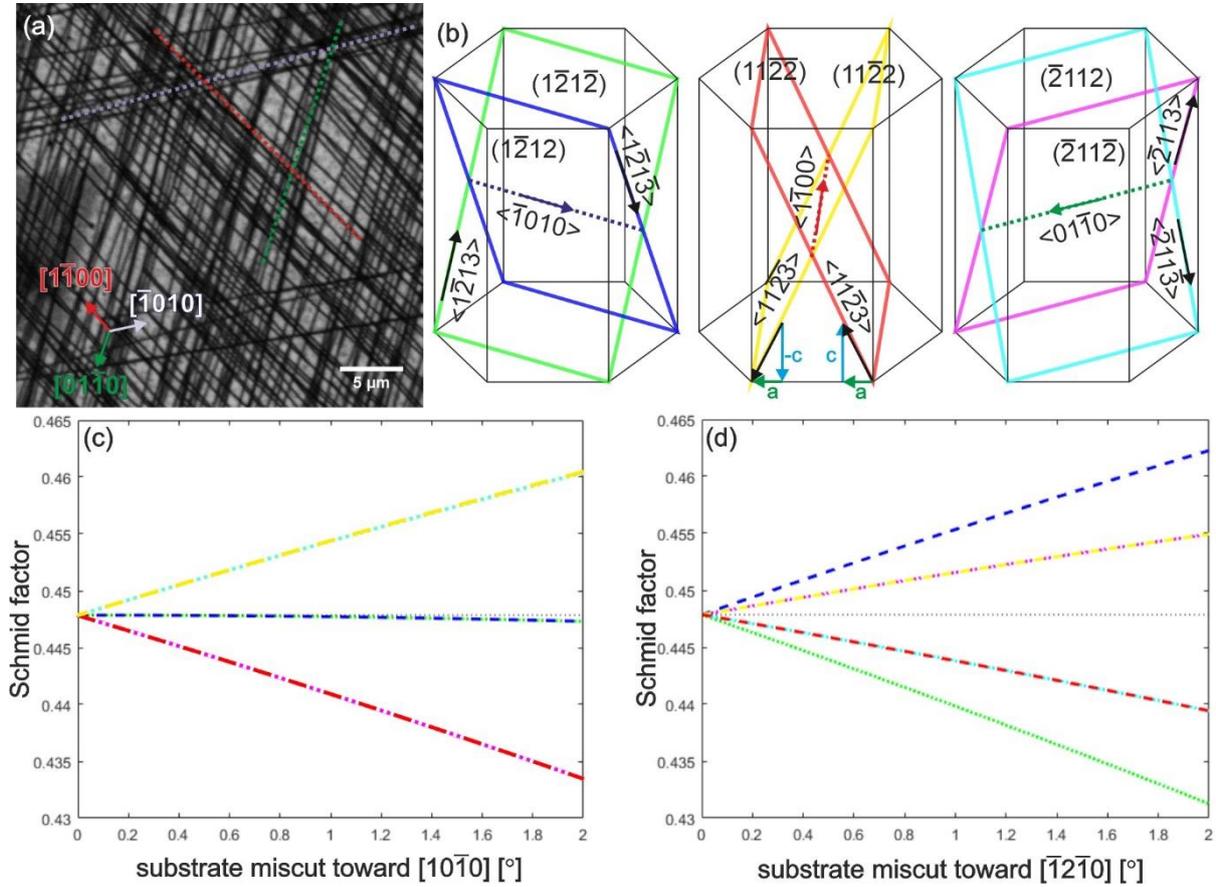

Fig. 2. (a) CL planar view image of a 50 nm thick $In_{0.2}Ga_{0.8}N$ layer. Misfit dislocations are visible as dark lines. The colors of the lines indicate the respective ⟨11,¯00⟩ directions. (b) Schematic drawing of the pairs of mirror-related {112,¯2} planes and the corresponding ⟨11,¯00⟩ directions. In-plane ('**a**') and out-of-plane ('**c**') components of ¹/₃[112,¯3] and ¹/₃[112,¯3,¯] Burgers vectors are marked with green and blue arrows, respectively. (c) – (d) Plots of calculated Schmid factors for each slip system as a function of the substrate misorientation. The colors of the lines correspond to the respective color-coded slip systems as shown in (b). The black dotted line indicates the value of the Schmid factor for the perfectly (0001)-oriented substrate.

Burgers vector, i.e., so-called (**a**+**c**) dislocations form a trigonal network along ⟨11,¯00⟩ directions at the InGaN/GaN interface (Fig. 2(a)). The alignment of misfit dislocations along ⟨11,¯00⟩ directions suggests dislocation nucleation and glide on the {112,¯2} slip planes (Fig.

2(b)). There are six {112̄2} crystallographic planes in the wurtzite structure accessible for nucleation and glide of (**a+c**) dislocations. The planes can be grouped into three pairs of mirror-related {112̄2} planes that share common ⟨11̄00⟩ directions e.g., the (12̄12) and the (12̄12̄) planes. Glide of a misfit dislocation half-loop on each of the mirror pairs results in one set of parallel misfit dislocations lying along the respective ⟨11̄00⟩ direction (Fig. 2(a)). Owing to the symmetry of the crystallographic system, the specific dislocation lying in a particular ⟨11̄00⟩ direction can have one of four possible Burgers vectors (Table 1). To relieve the misfit strain, all dislocations belonging to one set should have the same **a**-component. Dislocations from the common set may differ only in the **c**-component depending on the chosen mirror plane (Fig. 2(b)).

Table 1. {112̄2} slip planes with their corresponding Burgers vectors and resulting directions of the misfit dislocation (MD) line on the (0001) plane.

| Pyramidal plane | (2̄112̄) | (2̄112) | (12̄12̄) | (12̄12) | (112̄2̄) | (112̄2) |
|---|---|---|---|---|---|---|
| Burgers vector | ±⅓[2̄113] | ±⅓[2̄113̄] | ±⅓[12̄13] | ±⅓[12̄13̄] | ±⅓[112̄3] | ±⅓[112̄3̄] |
| MD line | [011̄0] | | [1̄010] | | [1̄100] | |

To analyze dislocation formation, the stresses acting in each slip system need to be determined. We transformed the stress tensor from the coordinate system related to the interface [σ] ((x, y, z), Fig. 1(c)) into a coordinate system related to the slip system [σ]″ ((x", y", z"), Fig. 1(c)) using transformation matrices defined in terms of Euler angles. Transformation details are presented in the Supplementary material.

In such coordinate system, the shear stress required for dislocation nucleation and glide is described by the $\sigma_{23}''$ stress component of the [σ]″ matrix.[*31*] In further analysis, we considered the Schmid factors of the respective slip systems. The Schmid factor *m* is a parameter describing a ratio between the resolved shear stress acting in the given slip system $\sigma_{23}''$ and the interfacial normal stress σ: $m = \frac{\sigma_{23}''}{\sigma}$. We calculated Schmid factors for six ⟨112̄3⟩{112̄2} slip systems for substrate misorientations toward ⟨1̄100⟩ and ⟨1̄1̄20⟩ directions. The results as a function of substrate misorientation are presented in Figs 2(c)-(d). According to the Schmid factor model, slip systems with the highest Schmid factors are favored for the introduction of misfit dislocations. For a layer deposited on a substrate with no misorientation the Schmid factor of each $\langle 11\bar{2}3\rangle\{11\bar{2}2\}$ slip system is the same and equal to m=0.4479. The same resolved shear stress acting on all $\langle 11\bar{2}3\rangle\{11\bar{2}2\}$ slip systems in the InGaN epilayer should lead to the formation of a regular trigonal dislocation network with the same dislocation density in each set as shown in Fig. 3(a). However, introducing the substrate misorientation changes the angles between the interface and each of the slip planes, thus modifying the resolved shear stresses in each of the slip systems.

In the case of substrate misorientation toward $[1\bar{0}10]$ (Fig. 2(c)), two sets of misfit dislocations along the $[01\bar{1}0]$ and $[1\bar{1}00]$ directions are expected to dominate, since dislocation formation is favored in the corresponding $\frac{1}{3}[2\bar{1}13](2\bar{1}\bar{1}2)$ (cyan in Fig. 2) and $\frac{1}{3}[11\bar{2}\bar{3}](11\bar{2}2)$ (yellow in Fig. 2) slip systems. Respective mirror-related slip systems would be less active due to lower Schmid factors than for the non-misoriented case. Misfit dislocations in the third set, along $[1\bar{0}10]$ direction, parallel to the substrate miscut, could be generated equally in both $\frac{1}{3}[1\bar{2}13](1\bar{2}1\bar{2})$ (green in Fig. 2) and $\frac{1}{3}[\bar{1}2\bar{1}3](\bar{1}2\bar{1}2)$ (blue in Fig. 2) mirror slip systems, but with lower nucleation rate due to higher activation energy (lower Schmid factor, close to non-misoriented case).

In turn, substrates misoriented toward **a**-direction would result in a different dislocation distribution. In the case of substrate misorientation toward $[1\bar{2}10]$ (Fig. 2(d)), one set of misfit dislocations along the $[1\bar{0}10]$ direction, perpendicular to the substrate miscut, is favored by the highest Schmid factor for the $\frac{1}{3}[\bar{1}2\bar{1}3](\bar{1}2\bar{1}2)$ slip system (blue in Fig. 2). In two other dislocation sets, i.e., $[01\bar{1}0]$ and $[1\bar{1}00]$ directions, the $\frac{1}{3}[2\bar{1}\bar{1}3](2\bar{1}\bar{1}\bar{2})$ (magenta in Fig. 3) and $\frac{1}{3}[11\bar{2}\bar{3}](11\bar{2}2)$ (yellow in Fig. 2) slip systems experience higher resolved shear stresses than the corresponding mirror planes. Schemes of dislocation networks expected for layers grown on misoriented substrates are shown in Figs 3(b)-(c). Fig. 3(d) illustrates expected in-plane deformation of the layer relaxed by irregular network of misfit dislocations.

Dislocation-induced layer-to-substrate tilt arises when the imbalance in dislocation generation with different out-of-plane Burgers vector component occurs as a result of the stress asymmetry. It has been observed in epitaxial systems crystallizing in cubic structures, like InGaAs/GaAs[*4,36*] or SiGe/Si[*10,37*] as well as in semipolar relaxed nitride structures[*14*]. The out-of-plane component of the Burgers vector introduces local tilt of the layer. In the case of partially plastically relaxed layers, the value of the resulting macroscopic layer-to-substrate tilt $\kappa$ is influenced by an interplay between the presence of the **c**-components of the misfit dislocations belonging to various sets as well as the coherency tilt $\kappa_n$ as described in part 2.1. [*18*].

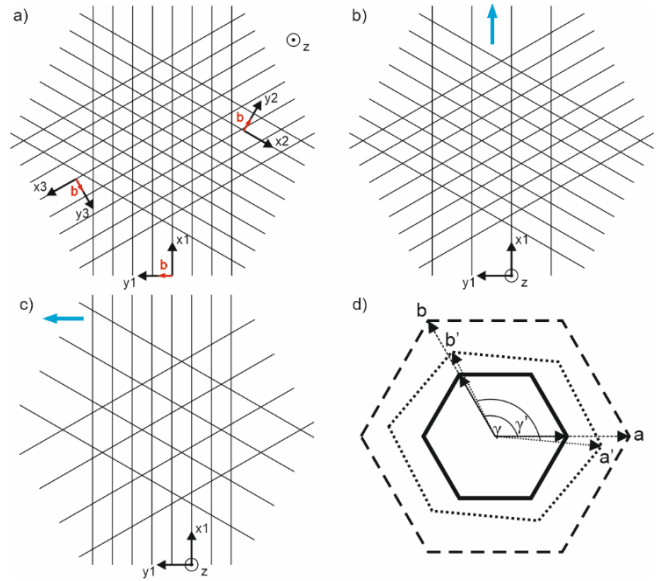

Fig. 3. Schemes of trigonal dislocation networks with (a) equal dislocation density in each set, (b) dislocation density twice lower in one set (corresponding to the dislocation distribution characteristic for InGaN layers grown on **m**-misoriented GaN substrates) and (c) dislocation density twice lower in two sets (corresponding to the dislocation distribution characteristic for InGaN layers grown on **a**-misoriented GaN substrates). Coordinate systems related to each dislocation sets, to the set with the lowest dislocation density and to the set with the highest dislocation density are marked in (a), (b) and (c), respectively. The red arrows indicate Burgers vectors. The blue arrow marks the misorientation azimuth. (d) Schematic illustration of the in-plane layer deformation induced by irregular misfit dislocation distribution. Solid and dashed lines correspond to wurtzite GaN and fully relaxed wurtzite InGaN, respectively. Then the lattice parameters a=b and γ=120° as well as $a_{GaN}$ is parallel to $a_{InGaN}$. Dotted lines correspond to partially relaxed InGaN where a'≠b' and γ'≠120°.

### 2.3. Strain state in a layer with irregular net of misfit dislocations

The irregularity of the dislocation densities in particular sets leads to an inequality of the strain tensor components. The strain relieved by a single array of misfit dislocations can be estimated as: $\delta = \frac{b_\parallel}{D}$, where $b_\parallel$ is an in-plane edge component of the Burgers vector and D is a distance between dislocations. Relaxation is most efficient perpendicular to the dislocation line. To estimate the total strain accommodated by the trigonal dislocation network one needs to add the strain components of all dislocation sets. The tensorial nature of the strain must be considered. In the coordinate systems related to the dislocation lines (Fig. 3(a)), the particular tensors of relieved strain for each dislocation set can be assumed to be $\delta = \begin{bmatrix} 0 & 0 & 0 \\ 0 & \delta & 0 \\ 0 & 0 & 0 \end{bmatrix}$. To add tensors one needs to transform them into the common coordinate system. The sum gives the total strain accommodated by the regular trigonal dislocation network (each dislocation set with $\delta = \frac{b_\parallel}{D}$):

$\delta_{reg} = \begin{bmatrix} \frac{3}{2}\delta & 0 & 0 \\ 0 & \frac{3}{2}\delta & 0 \\ 0 & 0 & 0 \end{bmatrix}$. This ensures isotropic relaxation.

The situation changes for irregular dislocation networks. For illustration, we considered a dislocation network where the misfit relieved in two sets is equal to $\delta$ and in the third set is twice less (i.e., $\frac{1}{2}\delta$), which corresponds to the dislocation distribution characteristic of InGaN layers grown on **m**-misoriented GaN substrates (Fig. 3(b)) as well as a network where misfit relieved in one set is equal to $\delta$ and in two others is twice less ($\frac{1}{2}\delta$), which corresponds to the dislocation distribution characteristic of InGaN layers grown on **a**-misoriented GaN substrates (Fig. 3(c)). Then, the total strain accommodated by the network is: $\delta_{tot}^m = \begin{bmatrix} \frac{3}{2}\delta & 0 & 0 \\ 0 & \delta & 0 \\ 0 & 0 & 0 \end{bmatrix}$ and

$\delta_{tot}^a = \begin{bmatrix} \frac{5}{4}\delta & 0 & 0 \\ 0 & \frac{3}{4}\delta & 0 \\ 0 & 0 & 0 \end{bmatrix}$, respectively.

The lattice misfit is expressed by $f=\varepsilon+\delta$,[31] where $\varepsilon$ is the residual elastic strain and $\delta$ is a part of the strain accommodated by misfit dislocations. Since $\delta_{xx} \neq \delta_{yy}$ and $\varepsilon = f - \delta$, then $\varepsilon_{xx}$ is no longer equal to $\varepsilon_{yy}$. Thus, an irregular trigonal misfit dislocation network leads to anisotropic relaxation, i.e., the in-plane lattice parameter of the layer and hence the degree of relaxation will be different for different crystallographic directions.

3. Experimental

$In_{0.2}Ga_{0.8}N$ layers were grown by Molecular Beam Epitaxy (MBE) at 630°C under metal-rich conditions. More details on the growth are provided in references [38,39]. Ammonothermal bulk GaN crystals of very low threading dislocation density ($10^4$ cm$^{-2}$) were used as substrates.

The first set consists of two samples with 50 nm thick layer grown on substrates with 1° misorientation toward ⟨112̄0⟩ direction (**a**-misoriented substrate) and 0.81° misorientation toward ⟨1̄100⟩ direction (**m**-misoriented substrate). Two sets of 100 nm thick layers were deposited in one growth process on substrates with various miscut angles: 0.23°, 0.76° and 1.42° misorientation toward ⟨112̄0⟩ direction (set A) and 0.27°, 0.81° and 1.01° misorientation toward ⟨1̄100⟩ direction (set M). The substrate misorientation was determined by XRD measurements with an accuracy of ±0.05°. The accuracy of the misorientation azimuth is of about a few degrees.

Structural and dislocation analysis was performed using transmission electron microscopy (TEM): FEI Tecnai G2 F20 S-TWIN operating at 200 kV and aberration-corrected FEI Titan 80-300 operating at 300 kV and by cathodoluminescence imaging (CL) using a Hitachi SU-70 SEM Microscope with a CLUE Jobin Yvon system attached. Cross sectional and planar specimens for TEM analysis were prepared by mechanical polishing followed by ion milling.

Layer tilt, chemical composition and strain state were established using the High-Resolution X-ray Diffractometry (HRXRD) employing reciprocal space mapping (RSM). It is common practice to assume hexagonal symmetry in XRD strain analysis of epitaxial nitride layers. There are then two unknown lattice parameters ($a$ and $c$) and measurements of only two Bragg reflections are required. Typically, parameter $c$ is found directly from a symmetric reflection (e.g., 0002 or 0004) and an asymmetric reflection is measured (e.g., 112̄4 or 101̄5) to determine parameter $a$, using the previously found value of $c$. Lattice parameter measurements provide information on the residual strain. Conventionally, the in-plane percent relaxation can be estimated by $R = \frac{a_L - a_S^0}{a_L^0 - a_S^0} \times 100$, where $a_L$ is the measured in-plane lattice parameter of the layer, $a_L^0$ is the stress-free in-plane lattice parameter of the layer and $a_S^0$ is the in-plane lattice parameter of the substrate.

However, if the lattice distortion is present, more measurements are required. The hexagonal symmetry of strained epitaxial wurtzite layers can only be assumed at very low strain levels. The triclinic lattice deformation implies the need for detailed XRD measurements to find six unknowns: $a$, $b$ and $c$ lattice parameters and $\alpha$, $\beta$ and $\gamma$ cell angles to accurately determine the chemical composition and the residual strain of the layer.

XRD measurements were performed using two high resolution diffractometers working with copper radiation (Cu$K_{\alpha 1}$, $\lambda$=1.5406 Å): Philips X'Pert MRD and Empyrean (Malvern Panalytical). Our approach consists of measuring six reciprocal space maps of 112̄4-type reflections and one 2θ-ω scan in tripple-axis mode of an 0002 symmetric reflection. Subsequent asymmetric maps were measured by rotating the sample clockwise every 60° around the normal to the surface. The position of an InGaN peak was determined as a mass center of the part of the peak that lies above the value ½$I_{max}$, where $I_{max}$ is the intensity of the highest point of the peak:

$$2\theta = \frac{\sum_n 2\theta_n(I_n - ½\, I_{max})}{\sum_n(I_n - ½\, I_{max})}.$$

Summation was performed over all points of the peak with intensity greater than ½$I_{max}$. $I_n$ and $2\theta_n$ are the intensity and $2\theta$ of the $n$-th point, respectively.

We determined the parameters *a, b, c,* α, β and γ of the InGaN lattice by iteratively fitting the measured 2θ angles of one (0002) symmetric and six $\{11\bar{2}4\}$ asymmetric reflections to the positions of the theoretical unit cell. Details of the calculations are presented in the Supplementary material. The direction of the found *a* parameter corresponds to the direction of the first measured asymmetric map. For the fully defined InGaN lattice, the tilt τ and its direction can be calculated from the lattice parameters using geometric relations. Next, both the strain and chemical composition must be determined together as both affect the lattice parameters. With a fully defined lattice, the strain and the composition can be found by the second iterative fitting. However, in the intuitive coordinate system where the **x**-axis is parallel to the **a**-axis of the unit cell and the **(x, y)** plane is parallel to the axes **a** and **b,** shear strain ε$_{xy}$ is not zero. Then, the degree of relaxation calculated with ε$_{xx}$ and ε$_{yy}$ components in such a coordinate system would be underestimated. To include the total strain in further considerations, we use tensor transformation to find the coordinate system **(x', y')** in which the shear strain ε$_{x'y'}$ is zero (ε$_{x'y'}$=0), i.e., the principal coordinate system. Such defined **x', y'** directions may deviate from the low indices crystallographic directions.

To illustrate a deviation from an in-plane symmetry, we define an anisotropy coefficient $A = \frac{\varepsilon_{y'y'}}{\varepsilon_{x'x'}}$, where ε$_{x'x'}$ and ε$_{y'y'}$ are calculated in-plane strain components defined in the coordinate system where ε$_{x'y'}$=0. Such a defined anisotropy coefficient A is equal to 1 for the isotropic layer and the deviation from this value increases with increasing anisotropy. We propose a way to express the degree of relaxation of partially relaxed wurtzite layers using the measured residual strain. The relaxation can then be described by the following two parameters: $R_{x'x'} = \frac{a_L^0 \varepsilon_{x'x'} + a_L^0 - a_S^0}{a_L^0 - a_S^0} \times 100$ and $R_{y'y'} = \frac{a_L^0 \varepsilon_{y'y'} + a_L^0 - a_S^0}{a_L^0 - a_S^0} \times 100$, where $a_L^0$ is the stress-free in-plane lattice parameter of the layer (the parameter of the unstrained cell calculated from Vegard's law for a given indium content), $a_S^0$ is the in-plane lattice parameter of the substrate and $\varepsilon_{x'x'}$ and $\varepsilon_{y'y'}$ are calculated in-plane strain components defined in the coordinate system for which $\varepsilon_{x'y'}=0$.

## 4. Results and discussion

TEM and CL analysis of InGaN layers reveal the trigonal network of (**a+c**)-type misfit dislocations aligned along $\langle 1\bar{1}00\rangle$ directions for all analyzed samples. Figs 4(a)-(b) show CL images of 50 nm thick $In_{0.2}Ga_{0.8}N$ layers exhibiting an initial relaxation state. The dominance of either two sets of dislocations (Fig. 4(a)) or one set (Fig. 4(b)) for **m**- and **a**-misoriented substrates, respectively, is clearly visible. The density of misfit dislocation lines in 100 nm $In_{0.2}Ga_{0.8}N$ layer grown on **m**-misoriented substrate is too high to visualize them in CL imaging, that is why we show the spatial distribution of dislocations in this layer using plan-view TEM image shown in Fig. 4(c). Dislocations are present in all three $\langle 1\bar{1}00\rangle$ directions indicating a higher degree of relaxation. However, it is still noticeable that the density of dislocations varies in different sets of dislocations.

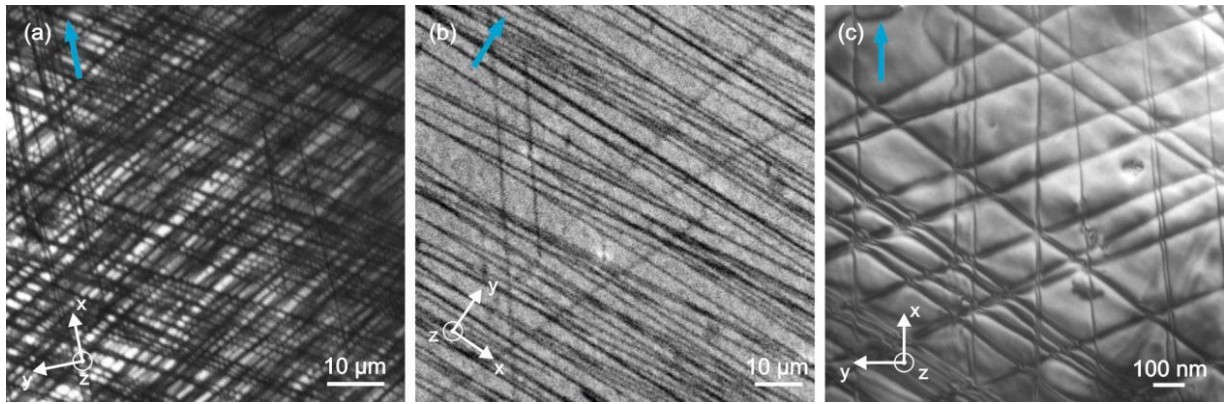

Fig. 4. CL planar view image of (a) 50 nm $In_{0.2}Ga_{0.8}N$ grown on **m**-misoriented substrate (0.81°) and (b) 50 nm $In_{0.2}Ga_{0.8}N$ grown on **a**-misoriented substrate (1°). (c) Plan-view bright-field TEM image of 100 nm $In_{0.2}Ga_{0.8}N$ grown on **m**-misoriented substrate (0.81°). Misfit dislocations are visible as dark lines. The blue arrow indicates the misorientation azimuth.

The estimated degrees of relaxation for 50 nm and 100 nm layers grown on **m**-misoriented substrate were calculated from the misfit dislocation density. Average distances between dislocations in particular sets were calculated from CL images (image area of 10x10μm) and TEM images (image area of 5x5μm) for 50 nm and 100 nm thick layers, respectively. The strain relieved by a particular dislocation ⟨1,$\bar{1}$00⟩ set was calculated as $\delta = \frac{b_\parallel}{D}$, where $b_\parallel$ is an in-plane edge component of the Burgers vector and D is an average distance between dislocations. Next, all three δ components were transformed into the common (x, y) coordinate system related to the set with the lowest dislocation density, as shown in Fig. 4. The tensor of total strain accommodated by the network was defined in this coordinate system by adding the components of all dislocation sets. The $\delta_{xx}$ and $\delta_{yy}$ components of the tensor of total strain accommodated by the network were used to estimate the respective degrees of relaxation: $R_{ii} = \frac{\delta_{ii}}{f}$, where f is the misfit. We found 6%/4% and 30%/20% of relaxation in x/y directions for 50 nm and 100 nm thick $In_{0.2}Ga_{0.8}N$ layers, respectively. The threading dislocation density in 50 and 100 nm layers is at the level of a few of $10^6$ and $10^8$ $cm^{-2}$, respectively. This is higher than the threading dislocation density of the substrate $10^4$ $cm^{-2}$, because each misfit dislocation half-loop introduces two threading segments.

We do not observe any clear correlation of the position of misfit dislocations with substrate atomic steps. The predicted spacing of the substrate steps changes from 64 nm for 0.23° miscut to 10 nm for 1.42° miscut. The average dislocation spacing is about 100 nm for 100 nm thick $In_{0.2}Ga_{0.8}N$. Since dislocations are formed by surface nucleation, it is likely that they nucleate at the surface steps but the experimental evidence for such nucleation is still lacking. It is worth noting that the InGaN surface during dislocation nucleation is very different from the perfect step morphology of the GaN substrate. Topographic analysis is shown in the Supplementary material. Our research so far shows a rather statistical dislocation distribution. Fig. 5 shows a high-resolution TEM image of two adjacent misfit dislocations with opposite **c**-components. A misfit dislocation has the full **c**-component of the Burgers vector and then introduces two (0002) extra half planes. Moreover, the core of misfit dislocations is dissociated with spatial distribution of partial dislocations along **a-** and **c-**directions as it was described elsewhere. [*40*] This means that its dimension is significantly larger than the step height of c/2.

The out-of-plane **c**-component of the dislocations introduces local tilt, visible as microscopic bending of the InGaN **c**-planes with respect to the GaN substrate (Fig. 5). To relieve the misfit strain, all dislocations belonging to one set should have the same in-plane **a**-component, however, no particular **c**-component is preferred in terms of strain relaxation. The out-of-plane **c**-component of their Burgers vectors, i.e., either "**+c**" or "**-c**", depends on the slip plane chosen from the mirror-pair planes. The structural analysis shows that the majority of dislocations lying along the particular direction have the same Burgers vector, especially the same **c**-component, as we also reported elsewhere[*35*]. The resulting macroscopic tilt of the entire layer κ (the angle between the (0001) planes of the layer and the (0001) planes of the substrate) depends on the total sum of the **c**-components of the misfit dislocations belonging to various sets.

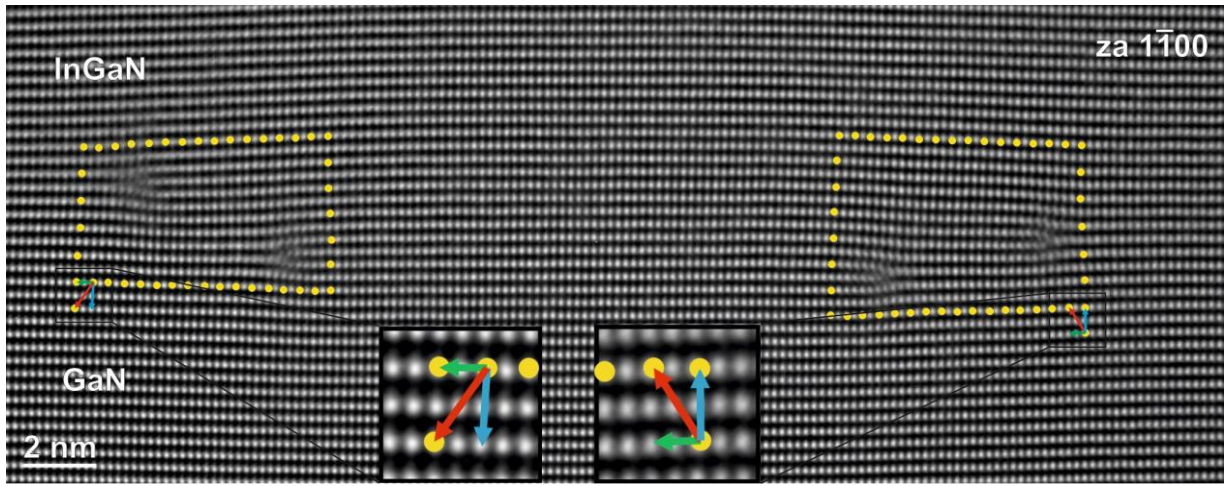

Fig. 5. High resolution TEM image taken along the [11$\bar{}$00] zone axis of two adjacent (**a**+**c**)-type misfit dislocations at the $In_{0.2}Ga_{0.8}N$/GaN interface laying on mirror related (112$\bar{}$2) planes. The projection of the Burgers vector onto the image plane determined by the Burgers circuit is marked with red arrows. The in-plane components of the Burgers vector (marked with green arrows) are the same for both dislocations, while the out-of-plane components of the Burgers vector (marked with blue arrows) are opposite. The bending of the **c**-planes in the InGaN lattice is noticeable.

The XRD measurements were performed for 100 nm InGaN layers of the series with variable substrate misorientation (A and M sets). As an example, Fig. 6 shows reciprocal space maps of six 112$\bar{}$4-type asymmetric reflections acquired for the InGaN layer deposited on 0.8° **m**-misoriented GaN (M2). Table 2 shows the determined lattice parameters and Table 3 the calculated results of triclinic deformation, In content, strain analysis and layer tilt.

The indium content was found to be close to the nominal value in all samples. Decreasing indium content with increasing substrate miscut is a known phenomenon for InGaN layers grown by Metalorganic Vapor Phase Epitaxy in the step-flow growth mode[*16*]. However, Molecular Beam Epitaxy technique employs a much lower growth temperature and MBE-grown In$_{0.2}$Ga$_{0.8}$N layers exhibit island formation on the (0001) surface which disrupts step-flow growth mode.[*38*] Then, it results in a weaker than expected dependence of indium incorporation as a function of substrate misorientation.[*38,41*]

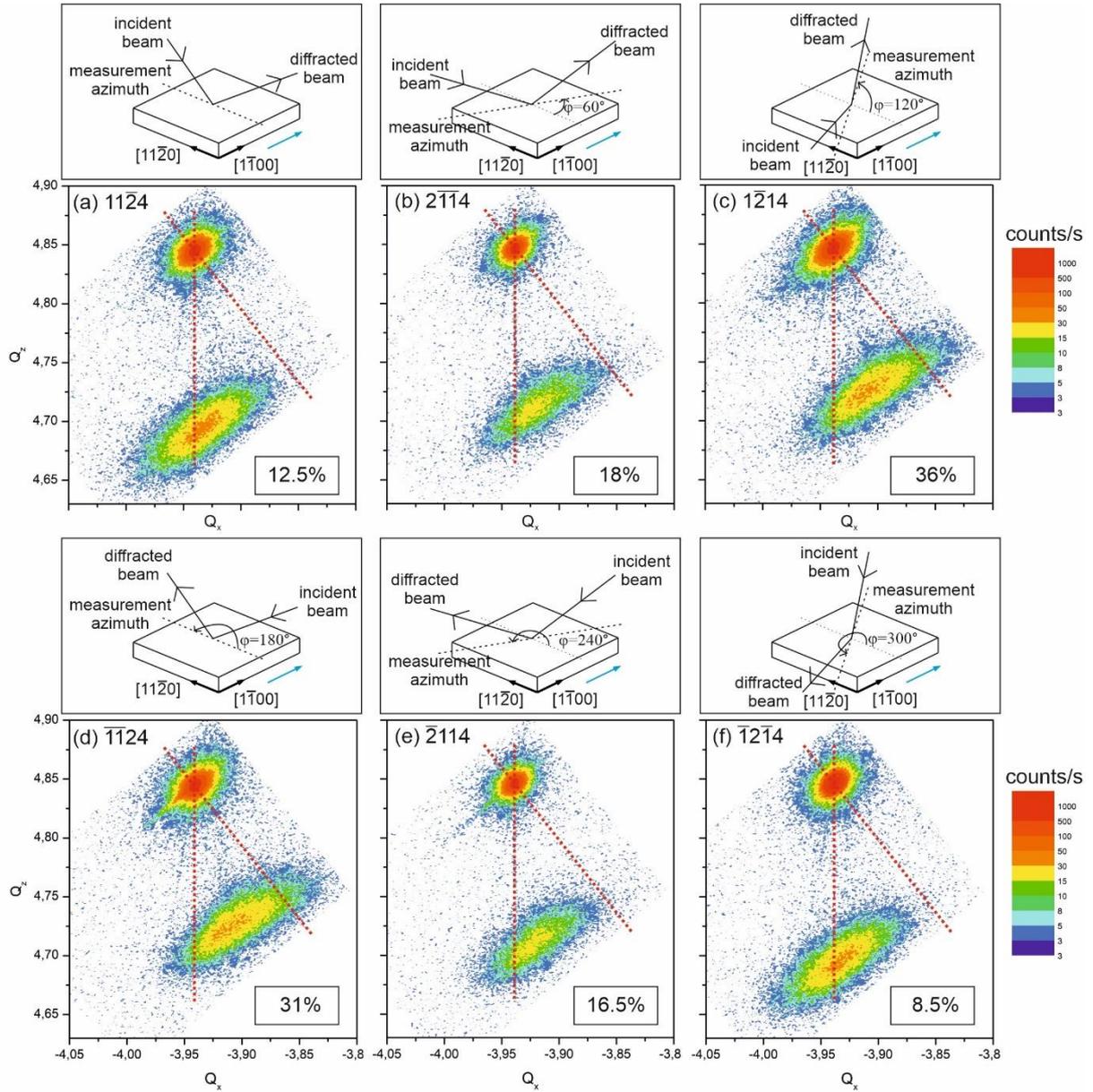

Fig. 6. Reciprocal space maps around asymmetric reflections a) 112$\bar{4}$, b) 21$\bar{1}\bar{4}$, c) 12$\bar{1}$4, d) 1$\bar{1}\bar{2}$4, e) 2$\bar{1}$14, f) 1$\bar{2}$1$\bar{4}$ for 100 nm thick In$_{0.2}$Ga$_{0.8}$N layer grown on 0.8° **m**-misoriented GaN (M2 sample) with measurement geometry. The blue arrow indicates the misorientation azimuth. The relaxation triangle is marked with red lines. The insets show the false degree of relaxation calculated for specific reflections assuming hexagonal symmetry (not triclinic one).

Table 2. XRD results of lattice parameters calculated for 100 nm InGaN layers. The estimation of the error bars is explained in the Supplementary material.

| Sample No. (substrate misorientation [°]) | a [Å] ±0.0014 | b [Å] ±0.0014 | c [Å] ±0.0008 | α [°] ±0.02 | β [°] ±0.02 | γ [°] ±0.03 |
|---|---|---|---|---|---|---|
| A1 (0.23) | 3.2068 | 3.2069 | 5.3345 | 90.02 | 90.01 | 119.99 |
| A2 (0.76) | 3.2058 | 3.2061 | 5.3323 | 90.04 | 90.03 | 119.91 |
| A3 (1.42) | 3.2021 | 3.2018 | 5.3253 | 90.07 | 90.07 | 119.86 |
| M1 (0.27) | 3.2056 | 3.2058 | 5.3331 | 90.03 | 89.97 | 120.01 |
| M2 (0.81) | 3.2075 | 3.2070 | 5.3321 | 90.06 | 89.93 | 120.03 |
| M3 (1.01) | 3.2043 | 3.2066 | 5.3279 | 90.09 | 89.91 | 120.04 |

Table 3. XRD results of In content, triclinic deformation, degree of relaxation and layer tilt for 100 nm InGaN layers. The estimation of the error bars is explained in the Supplementary material.

| Sample No. (substrate misorientation [°]) | In content ±1 [%] | anisotropy coefficient $A = \frac{\varepsilon_{y'y'}}{\varepsilon_{x'x'}}$ | deformation tilt τ ±0.015[°] (the value predicted by the elasticity theory) | Rx'x' ±3 [%] | Ry'y' ±3 [%] | layer tilt κ ±0.02 [°] |
|---|---|---|---|---|---|---|
| A1 (0.23) | 19 | 0.99 | 0.035 (0.02) | 26 | 27 | 0.02 |
| A2 (0.76) | 19 | 0.88 | 0.066 (0.06) | 23 | 32 | 0.15 |
| A3 (1.42) | 18 | 0.83 | 0.133 (0.11) | 17 | 31 | 0.27 |
| M1 (0.27) | 19 | 0.99 | 0.033 (0.02) | 24 | 25 | 0.04 |
| M2 (0.81) | 19 | 0.97 | 0.073 (0.07) | 25 | 28 | 0.17 |
| M3 (1.01) | 18 | 0.93 | 0.106 (0.08) | 22 | 27 | 0.22 |

The triclinic lattice distortion is notable for layers grown on misoriented GaN substrates, although the differences in the lattice parameters are in the range of measurement accuracy (Table 2). It was found that the tilt τ is opposite to the miscut direction and it increases with increasing misorientation, evidencing a more pronounced distortion (Table 3). The values are slightly higher than expected from the elasticity theory for fully coherent layers indicating impact of misfit dislocations on triclinic distortion. It is accompanied by a more pronounced in-plane anisotropy, which also increases with increasing misorientation: the anisotropy coefficient $A = \frac{\varepsilon_{y'y'}}{\varepsilon_{x'x'}}$ deviates more from the value of one. We found out that the observed relaxation mechanism is very sensitive to substrate misorientation. The lowest anisotropy occurs for the low miscut angles. Layers grown on 0.3° misoriented substrates show almost no anisotropy in the in-plane relaxation. Substrate misorientation of about 0.8° is large enough to introduce significant differences in degrees of relaxation. The largest difference is found for the A3 sample, where the relaxation changes from 17% to 31%, so almost twice, for orthogonal directions. Substrate misorientation toward **a**-direction introduces stronger anisotropy than the corresponding misorientation toward **m**-direction. Meanwhile, the average degree of relaxation for the corresponding layers is very similar. The degree of relaxation is comparable to the values obtained from the dislocation density (measured on plan-view CL and TEM images Fig.4), indicating that the proposed $R_{x'x'}$ and $R_{y'y'}$ parameters are reliable.

It is a common practice in the literature to present a reciprocal space map of only one asymmetric reflection of relaxed InGaN. The insets in Fig. 6 show the degree of relaxation calculated according to the standard procedure (hexagonal symmetry approximation) for all measured reflections with correction for layer tilt included. The values obtained with the hexagonal symmetry approximation differ significantly from each other and from the values calculated for the triclinic lattice. Since we have evidenced anisotropy in the dislocation distribution, we expect different degrees of relaxation in all three **a**-directions. However, the obtained values varied significantly even after 180° specimen rotation, e.g., from 8.5% to 36% for $1 2,\bar{1} 4$ and $1,\bar{2} 1,\bar{4}$ reflections, respectively. In this case, equivalent results should be obtained since one is probing the same crystallographic direction. This makes such measurements unreliable. It appears that the hexagonal symmetry approximation is invalid and the standard procedure leads to significant errors in the study of relaxed wurtzite epitaxial layers on the miscut substrates. The presentation of just a single reciprocal space map can then give misleading information about the relaxation state.

The respective tilt κ between InGaN and GaN (0001)-planes was also measured by the XRD. For 50 nm thick layers, the dislocation density is too low to introduce a significant macroscopic tilt. It is about 0.02°, which is close to the estimated Nagai's tilt for this layer (0.016°). This layer is at the very beginning of the plastic relaxation process and it is nearly coherent with the substrate. The tilt increases as layer thickness and degree of relaxation increase. Table 3 presents results for 100 nm thick layers. The tilt κ increases with increasing miscut and becomes significant with respect to the substrate misorientations used, e.g., the tilt of the M2 layer is 0.17° with respect to a substrate misorientation of 0.81°. This is by an order of magnitude larger than the Nagai's tilt expected for coherent layers. The Nagai's tilt contributes to the total layer-to-substrate tilt κ, however, it decreases with increasing degree of plastic relaxation and tends to zero for fully relaxed layers. The measured tilt κ can be then explained by the preferential formation of misfit dislocations with a specific out-of-plane component due to the stress asymmetry introduced by the substrate misorientation. It was found that the layer-to-substrate tilt is opposite to the miscut direction, which leads to the decrease of the misorientation of the overall InGaN/GaN structure.

The crystal curvature was measured in two orthogonal directions: toward and perpendicular to the substrate miscut. We found a small bowing radius (a few meters) of the InGaN layers while the underlying GaN substrates remain relatively flat. We continue investigation to understand this effect, which likely originates from the specificity of misfit dislocations and strong lattice deformation.

## 5. Conclusions

In the case of InGaN layers deposited on misoriented substrates, not all $\langle 11\bar{2}3\rangle\{11\bar{2}2\}$ slip systems are equivalent for (**a**+**c**)-dislocation nucleation and glide which follows from the observation of different densities of misfit dislocations along different $\langle 1\bar{1}00\rangle$ directions in CL and TEM studies and the differences in the density of dislocations with opposite **c**-components. Both of these effects can be qualitatively explained by the Schmid factor model and attributed to the changes in resolved shear stresses induced by substrate misorientation. According to the model, the misorientation of a GaN substrate toward **a**- or **m**-directions in the wurtzite lattice result in the dominance of one or two sets of misfit dislocations, respectively. Such a mismatch in the dislocation distribution leads to strain anisotropy in the InGaN epitaxial layers, which is confirmed by our structural studies. As expected from the dislocation distribution model, it was found that higher misorientation decreases the energy for dislocation nucleation in some glide planes, while it has an opposite effect in others, leading to higher strain anisotropy. The layer tilt also increases as the substrate miscut increases due to enhanced imbalance in dislocation formation on the mirror-related $\{11\bar{2}2\}$ planes. It is worth noting that the tilt of the layer with respect to the substrate reduces the initial substrate misorientation resulting in a different total misorientation of the final structure. Since substrate misorientation strongly affects the properties of InGaN layers[*16*], the InGaN tilt should also be considered in the design of such substrates for the potential application of relaxed InGaN as pseudo-substrates. Layers grown on 0.3° misoriented substrates show almost no anisotropy in the in-plane relaxation along with relatively small layer tilt, however, layers grown on 0.8° misoriented substrates exhibit significant lattice deformation.

We observed the discussed phenomena in a wide range of plastically relaxed InGaN layers grown by MBE or MOVPE techniques. The observed mechanism is universal, independent of the type of GaN substrate and its threading dislocation density (TDD), such as bulk GaN substrates prepared by halide vapor phase epitaxy (TDD ~$10^6$ cm$^{-2}$) or by ammonothermal growth (TDD ~$10^4$ cm$^{-2}$) or on GaN/sapphire templates (TDD ~$10^8$ cm$^{-2}$).

In summary, we have demonstrated that InGaN layers grown on misoriented substrates relax by preferential activation of certain glide planes for misfit dislocation formation. We observe two different aspects of preferential dislocation formation: (i) the preferential glide on one plane from mirror related pairs leading to the tilting of the layer with respect to the substrate and (ii) the preferential formation of dislocations in particular $\langle 1\bar{1}00\rangle$ sets leading to in-plane anisotropy of the layer. Such an imbalance in dislocation generation is the result of the GaN substrate misorientation, which changes the shear stresses acting in the respective glide planes, resulting in a different dislocation nucleation ratio in each dislocation set. Accordingly, the perfect wurtzite symmetry of the layer is no longer preserved and triclinic deformation of the lattice is observed. This implies the need for detailed XRD measurements of the strain state of the layer. The estimation of the degree of relaxation from the single asymmetric Bragg reflection, which is a common practice in the literature, can then lead to misleading conclusions. We observe substantial differences in the estimated relaxation state depending on the chosen $\{11\bar{2}4\}$ reflection. It appears that to properly determine the properties of relaxed InGaN by XRD measurements, it is necessary to include the triclinic distortion of the InGaN lattice. We defined the InGaN lattice by measuring the positions of a symmetric and all six $\{11\bar{2}4\}$ asymmetric reflections. We propose to describe the strain relaxation of anisotropic partially relaxed InGaN layers by two parameters of degree of relaxation: $R_{x'x'}$ and $R_{y'y'}$ defined by the measured residual strains in two orthogonal directions found in a coordinate system for which the in-plane shear strain is zero. Moreover, preferential dislocation formation introduces additional tilt of the layer with respect to the substrate which changes the total misorientation of the final structure. These features would influence the properties of the layers deposited on such plastically relaxed InGaN. Thus, the development of structures grown on relaxed InGaN buffers should take into account these effects.

**Acknowledgments**

This work was funded in part by National Science Center, Poland, project OPUS LAP 2020/39/I/ST5/03379 and by Deutsche Forschungsgemeinschaft, Germany, project 465219948. For the purpose of Open Access, the author has applied a CC-BY public copyright license to any Author Accepted Manuscript (AAM) version arising from this submission. This research was also supported by Project of National Science Center 2018/31/G/ST5/03765.